\documentclass{PoS}

\usepackage{graphicx}

\title{Ultraluminous X-ray sources}

\ShortTitle{Ultraluminous X-ray sources}

\author{\speaker{S. Fabrika}
\\
        Special Astrophysical Observatory, Russia\\
        E-mail: \email{fabrika@sao.ru}}

\abstract{
The origin of Ultraluminous X-ray Sources (ULXs) in external galaxies whose X-ray luminosities exceed those of the brightest black holes in our Galaxy by hundreds and thousands of times is mysterious. The most popular models for the ULXs involve either intermediate mass black holes (IMBHs) or stellar-mass black holes accreting at super-Eddington rates. Here we review the ULX properties. Their X-ray spectra indicate a presence of hot winds in their accretion disks supposing the supercritical accretion. In recent years, new surprising results were discovered in X-ray data, ULX-pulsars and high-velocity outflows up to 0.2\,c. They are also in accordance with the super-Eddington accretion. However, the strongest evidences come from optical spectroscopy. The spectra of the ULX counterparts are very similar to that of SS433, the only known supercritical accretor in our Galaxy. The spectra are apparently of WNL type (late nitrogen Wolf-Rayet stars) or LBV (luminous blue variables) in their hot state, which are very scarce stellar objects. We find that the spectra do not originate from WNL/LBV type donors but from very hot winds from the accretion disks, which have similar physical conditions as the stellar winds from these stars. The results suggest that bona-fide ULXs must constitute a homogeneous class of objects, which most likely have supercritical accretion disks.
}

\FullConference{Accretion Processes in Cosmic Sources -- APCS2016 --\\
		5-10 September 2016,\\
		Saint Petersburg, Russia}

\begin{document}

\section{Introduction}

Ultraluminous X-ray sources (ULXs) are X-ray sources with luminosities
exceeding the Eddington limit for a typical stellar-mass black hole
$\sim 2 \times 10^{39}$ erg s$^{-1}$. Despite their importance
in understanding the origin of supermassive black holes that reside
in most of present galaxies, the basic nature of ULXs remains unsolved 
\protect{\cite{feng11,Bachetti2016}}. The most popular
models for the ULXs involve either intermediate mass black holes (IMBH,
$10^3$--$10^4$\,M$_{\odot}$) \protect{\cite{madau01}} with standard accretion disks or
stellar-mass black holes ($\sim 10$\,M$_{\odot}$) accreting at super-Eddington
rates. The last idea has been suggested \protect{\cite{fabr01}} because of an analogy
with SS\,433, the only known super-accretor in the Galaxy \protect{\cite{fabr1997,fabr04}}. 
It was proposed that SS\,433 supercritical disk's funnel being observed nearly 
face-on will appear as extremely bright X-ray source. In ULXs a moderate beaming in
the funnel is needed to produce the super-Eddington X-ray luminosities.
Both scenarios, however, require a massive donor in a close binary.

In last years the community has accepted a concept of the super-Eddington accretion 
with stellar mass black holes. It was not only related with finding of ultraluminous 
X-ray pulsars \protect{\cite{Bachetti2014,Furst2017,Israel2017}}, there were many new 
interesting discoveries. At the moment three ULX pulsars were found, it was 
a notable achievement. All the pulsars have their X-ray spectra harder than those average 
in ULXs. Many models have been suggested for the ULX-pulsars including magnetic field 
of different strength. However, one of the best model is probably the same geometrical beaming
\protect{\cite{Kawashima2016}} to violate the Eddington limit.

The X-ray spectra of the ULXs often show a high-energy curvature with
a downturn between $\sim 4$ and $\sim 7$\,keV \protect{\cite{Stobbart2006}}. It was called ``ultraliminous
state'' \protect{\cite{gladst09}}. The curvature hints that the ULX accretion disks
are not standard. Inner parts of the disks may be covered with hot outflow
or optically thick corona, which Comptonizes the inner disk photons. In more detail study
\protect{\cite{Sutton2013}} it was suggested an empirical classification scheme of the ULXs:
a singly-peaked broadened disk with a maximum to harder band (BD), and two spectral components
as hard ultraluminous (HUL) and soft ultraluminous (SUL) regimes. One more class of the sources
called extremely soft ultraluminous regime (ULS) \protect{\cite{Urquhart2016}}. In some nearby regimes the 
sources might jump between BD--HUL, HUL--SUL \protect{\cite{Sutton2013}} or SUL--ULS \protect{\cite{Pinto2016a}}.

Although all the previous X-ray data could not prove reliably the super-Eddington accretion, 
recently such evidences have been found. In several ULXs high velocity outflows have been detected
in the X-ray spectra. They are blueshifted absorption lines with velocities about 0.2\,c arising in highly ionized gas. Originally such absorption lines and edges have been predicted for ULXs in 
\protect{\cite{Fabrika2006,Fabrika2007}} on the grounds of SS\,433. 
They are common residuals which depend however on the outflow
velocity. The residuals have been clearly suspected \protect{\cite{Middleton2015}} in several ULXs and have
been discovered in \protect{\cite{Pinto2016b}} (NGC1313 X-1 and NGC5408 X-1), in \protect{\cite{Soria2016}} (M101 ULX-1), 
and in \protect{\cite{Pinto2016a}} (NGC55 ULX). The blueshifted absorption lines do not depend on the 
spectral components, in NGC1313 X-1 the regime is HUL, in NGC5408 X-1 - SUL, in NGC55 ULX - SUL/ULS, and in M101 ULX-1 - ULS. 

Most of the ULXs are associated with the star-forming regions and surrounded
with nebulae of a complex shape, indicating a dynamical influence of the
black hole \protect{\cite{abol07,Grise2012,Cseh2012}}. In Holmberg II X-1 there was a direct detection 
\protect{\cite{Lehmann2005,Egorov2017}} of nebula surrounded the ULX with a velocity dispersion 
of several tens of km/sec. That may be related with jets or strong outflows.

In a case of IMBH they are not distributed throughout galaxies as it
would be expected for IMBHs originating from low-metallicity Population
III stars \protect{\cite{madau01}}. The IMBHs may be produced in a runaway merging in a core of
young clusters. In this case, they should stay within the clusters. It has been 
found \protect{\cite{pout13}} that all brightest X-ray sources in Antennae galaxies are 
located nearby to very young stellar clusters with an age less than 
5 Myrs. It was concluded that the sources were ejected in the
process of formation of stellar clusters in the dynamical few-body
encounters. An average velocity of the ULXs in Antennae is $\sim 80$\,km/sec. 
Therefore, the majority of ULXs are massive X-ray binaries with the progenitor masses 
larger than 50\,M$_{\odot}$.

However the most obvious evidence that the ULXs are super-Edington accretion disks
came from optical spectroscopy \protect{\cite{fabr15}}. It was found that spectra of ULXs
are very similar to SS\,433 \protect{\cite{kubota10}} and LBV stars in their hot state \protect{\cite{shol11}}
or WNLh stars \protect{\cite{crowther11}}. All these objects have strong and dense outflows.

\begin{figure}
\hspace{1.1cm}
\includegraphics[width=.8\textwidth]{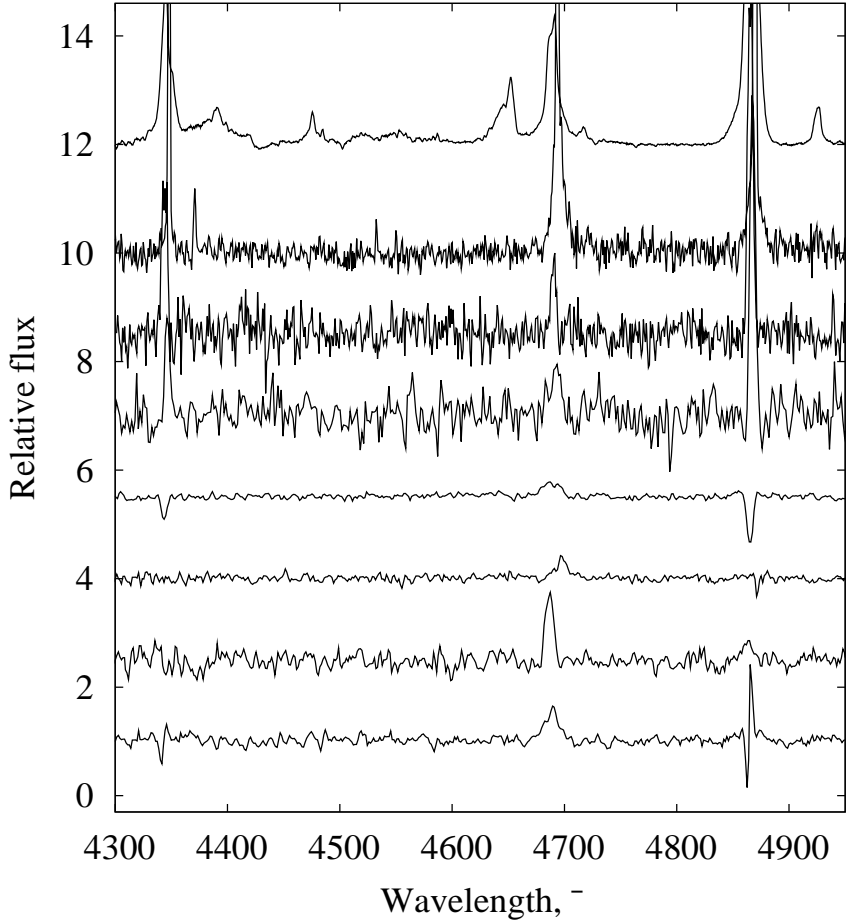}
\caption{Normalized optical spectra of ULX counterparts. From top to bottom: SS\,433 (1),
NGC\,5408 X-1 (2), NGC\,4395 X-1 (3), NGC\,1313 X-2 (2), NGC\,5204 X-1 (1), NGC\,4559 X-7 (1),
Holmberg\,IX X-1 (1) and Holmberg\,II X-1 (1). The numbers in brackets mean optical telescopes:
1 -- Subaru telescope, 2 -- VLT (ESO), 3 -- Russian BTA telescope. The spectra are very
similar one another, they may represent rare type of massive stars WNLh \protect{\cite{crowther11}}
or LBV stars in their hot states \protect{\cite{shol11}}. All the spectra are also similar
to SS\,433 \protect{\cite{kubota10}}. This means that the spectra of the ULX counterparts are formed
in hot winds.}
\label{fig1}
\end{figure}

\section{Optical spectra of ULXs}

Optical spectra of almost all ULX counterparts (with SS\,433 included)
are shown in Fig.\,1. Other spectra not shown here published in \protect{\cite{Motch2014}}
(NGC7793 P13, ULX-pulsar) and in \protect{\cite{Liu2015}} (M81 ULS-1).
Main features in all the spectra are the bright He\,II\,$\lambda 4686$, hydrogen
H$\alpha$, and H$\beta$ emission lines. The lines are obviously broad, the widths
range from 500 to 1500\,km s$^{-1}$.

The spectra extraction was not simple due to the nebular emission that
surrounds almost all the ULX counterparts. We extracted the spectra
with Gauss-profile aperture, and made two versions of the spectra. In
the first one, 
we used extended regions of 4--10 arcsec for the background.
Although this gives a higher S/N in the final spectra, the hydrogen 
lines originating from the nebulae may strongly contaminate the ULX
spectrum. In the second version,  
we used very small regions for the background, $\approx$1 arcsec, and
produced the spectra with the least contamination by the nebular
lines. 

\begin{figure}
\hspace{1cm}
\includegraphics[width=.8\textwidth]{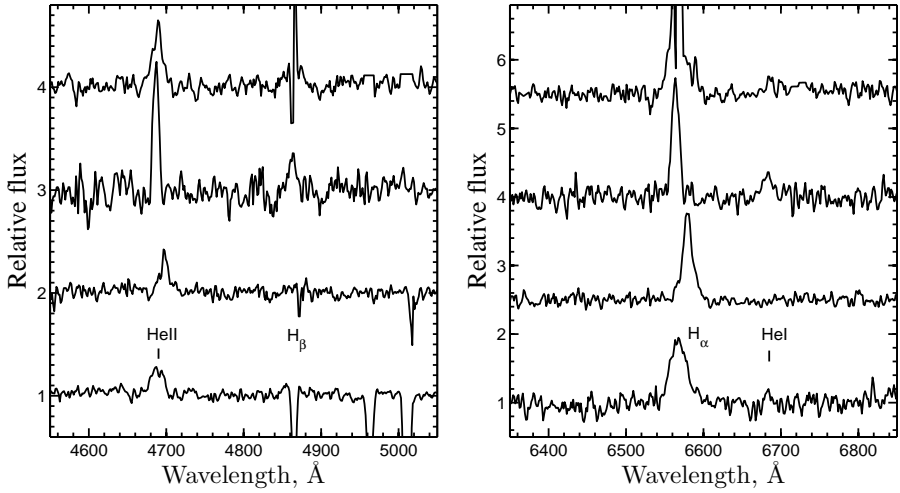}
\caption{ Spectra of the ULX optical counterparts from top to
bottom, Holmberg\,II, Holmberg\,IX, NGC\,4559, and NGC\,5204 in blue
({\bf a}) and red ({\bf b})
spectral regions. The spectra are normalized for better inspection.
The most strong are the He\,II $\lambda 4686$ line and the hydrogen lines
H$\alpha$ $\lambda 6563$ and H$\beta$ $\lambda 4861$.
The broad He\,I\,$\lambda 6678$ line is also detected. Narrow nebular emission
in H$\beta$ and [O\,III] $\lambda \lambda 4959, 5007$ lines is oversubstacted.
Although the hydrogen lines are contaminated with the nebular emission,
their broad wings are clearly seen.}
\label{fig2}
\end{figure}

We produced the normalized spectra for better inspection of spectral
features (Fig.\,2) taken with the
Subaru telescope \protect{\cite{fabr15}}. The main feature in all the spectra is a broad
He\,II\,$\lambda 4686$ emission line.  The most narrow one is found from
Holmberg\,IX, with an FWHM\,$\approx 450$\, km s$^{-1}$, and the most
broad one is from NGC\,5204, with an FWHM\,$\approx 1570$\,km s$^{-1}$. 
All line widths are corrected for the spectral resolution.

Calibrated spectra of the ULX optical counterparts are given in Fig.\,3
(the same as in Fig.\,2).
We conclude that all ULX counterparts ever spectrally observed have the
same feature in their spectra, that is, a broad He\,II emission line. We
also clearly detect broad H$\alpha$, H$\beta$ lines and
He\,I\,$\lambda 6678, 5876$ lines (Fig.\,3). There is also some
hints on the Bowen C\,III/N\,III blend (4640 - 4650~\AA). Although the
H$\beta$ line (Fig.\,2, 3) is affected by nebular emission in spite of our careful
extraction, its broad wings are clearly detected. It is obvious
that the emission lines are formed in stellar winds or disk winds.

\begin{figure}
\hspace{1cm}
\includegraphics[width=.8\textwidth]{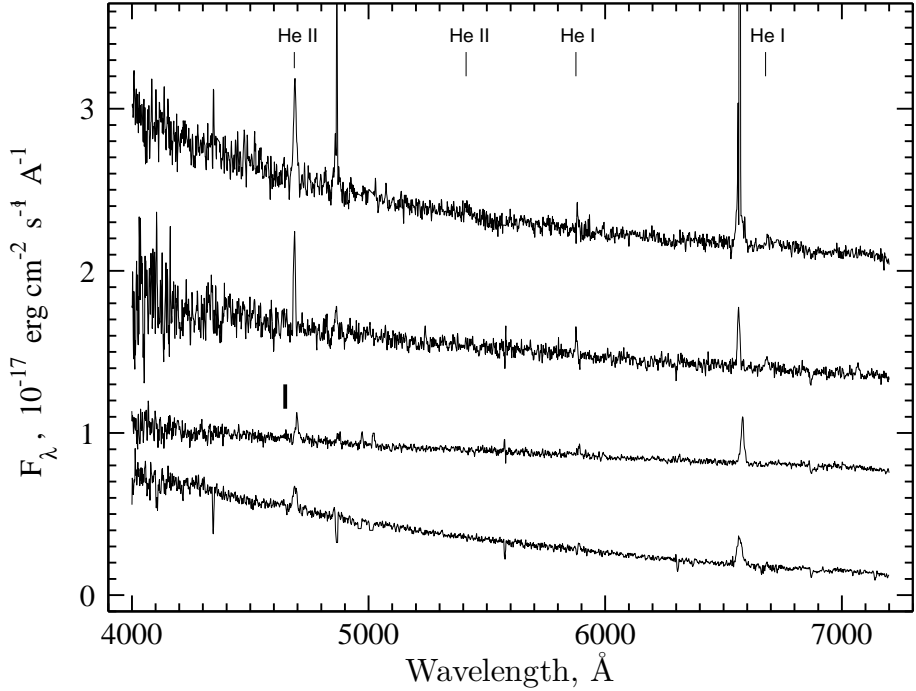}
\caption{Calibrated spectra of the ULX optical counterparts taken with the
Subaru telescope.
From top to bottom: the ULX in Holmberg\,II, Holmberg\,IX, NGC\,4559,
and NGC\,5204. For better visualization we add flux offsets of 1.8, 1.2 and 0.6
($10^{-17}$\,erg/cm$^2$s\,\AA~) for the Holmberg\,II, Holmberg\,IX, and
NGC\,4559 ULXs, respectively. Besides obviuos hydrogen lines
we mark He\,II lines ($\lambda 4686$ and $\lambda 5412$) and
He\,I lines ($\lambda 5876$ and $\lambda 6678$), thick bar indicates position of
the Bowen blend C\,III/N\,III $\lambda \lambda 4640 - 4650$.}
\label{fig3}
\end{figure}
                 
All studied ULX have broad He\,II\,$\lambda 4686$ and H$\alpha$. However, 
neither very hot WNLh nor LBV examples exhibit such strong He\,II emission 
lines relative to the hydrogen lines. If the abundance of hydrogen in the ULX 
donors were two times smaller from the Solar value, then it could make the He/H
ratio five times larger. This contradicts with the non-enhancement of 
the He\,I and Pickering He\,II lines indicating nearly normal abundance of hydrogen. 
Hence, the wind in the ULXs must be even hotter and more highly ionized
than stellar winds in WNLh or LBV stars.

All the spectra of the ULXs are surprisingly similar to one another.
The optical spectra are also similar to that of SS 433, although the ULX spectra
indicate a higher wind temperature. It was suggested in \protect{\cite{fabr15}} that
the ULXs must constitute a homogeneous class of objects, which most likely
have supercritical accretion disks.

Among stellar spectra, such a strong He\,II line with a nearly normal
hydrogen abundance can be found only in stars recently classified as
O2--3.5If/WN5--7 \protect{\cite{crowther11}}. 
They are the hottest transition stars, whose classification is based on
the H$\beta$ profile, tracing the increasing wind density (i.e., the mass loss rate)
from O2--3.5If, O2--3.5If/WN5--7, and to WN5--7.

\begin{figure}
\includegraphics[width=1.0\textwidth]{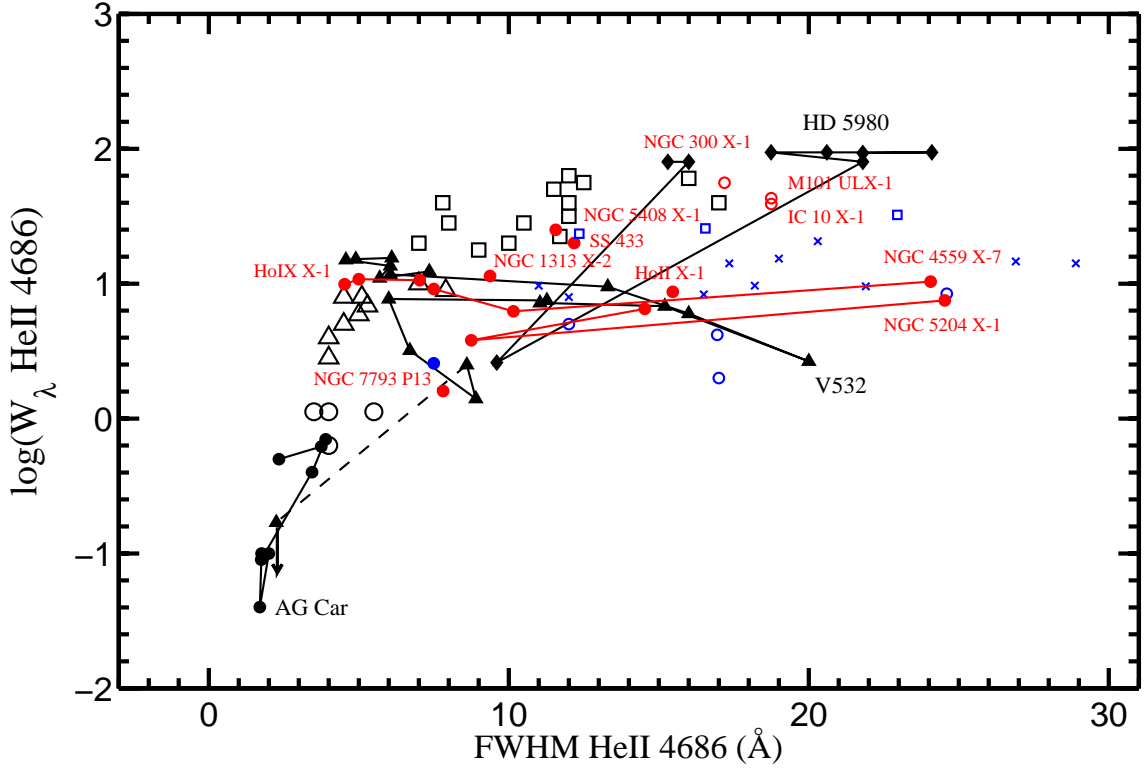}
\caption{Classification diagram of WNL stars in the LMC and our Galaxy \protect{\cite{Crowther1997}}.
The black open squares, triangles, and circles mark
WN\,8, WN\,9--10, and WN\,11 stars, respectively.
The blue filled circle denotes $\zeta$\,Pup. Other Galactic and LMC
stars \protect{\cite{crowther11}} are O2If and O3If (open blue circles), O2If/WN5, O2.5If/WN6,
O3If/WN6, and O3.5If/WN7 (blue crosses), and WN6ha and WN7ha stars (open blue
squares). There are three known LBV\,--\,WNL transitions (AG\,Car, V\,532,
and HD\,5980) in this diagram \protect{\cite{shol11}}. Consequent states of each
LBV star are connected with the lines.
Positions of our four ULX counterparts are also shown in red (connected with
lines to show variability from night to night), together with those of
NGC\,1313\,X-2, NGC\,5408\,X-1, NGC\,7793\,P13, SS\,433, NGC\,300\,X-1,
M\,101\,ULX-1, and IC\,10\,X-1 
\protect{\cite{roberts11,Cseh2011,Motch2014,kubota10,Crowther2010,kuntz05,silverman08}}.
}             
\label{fig4}
\end{figure}

In Fig.\,4 we show the classification diagram of WN stars \protect{\cite{Crowther1997}}
for LMC and Galactic objects. We supplement the diagram with additional
stars recently classified \protect{\cite{crowther11}}. The diagram plots stars in
accordance with their wind velocity (FWHM, the terminal velocity of a 
stellar wind), while the equivalent width (EW) reflects its photosphere 
temperature and a mass loss rate.
Three known LBV transitions (LBV\,--\,WNL) between their hot and cool
states in AG\,Car, V\,532 in M\,33, and HD\,5980 in SMC are also shown in
the figure. The transitions last from months to years. 
Consequent states in each LBV transition are connected with
the lines. In their hotter LBV-state where the He\,II line becomes stronger,
the LBVs fit well the classical WNL stars \protect{\cite{shol11}}.

In the figure, we also present two recently discovered extragalactic
black holes NGC\,300\,X-1 and IC\,10\,X-1 together with the soft ULX
transient M\,101\,ULX-1. The black holes in NGC\,300\,X-1 and IC\,10 have
luminosities $L_{\rm X} \sim 3 \times 10^{38}$\,erg s$^{-1}$, they contain the WN donors.
About the same as that of Cyg\,X-3, which certainly contains a WN-type donor
star \protect{\cite{vanKekrwijk96}}. The comparable luminosities with that of Cyg\,X-3, short orbital
periods, and the location in the diagram around the WN6--7 region confirm
that their optical spectra come from WN donors. The same may be proposed
for M\,101\,ULX-1 on the basis of its location in the diagram. It has
been recently found that this source indeed contains a WN8 type
donor \protect{\cite{Liu2013}}, although its orbital period is $\sim 40$ times longer than
in Cyg\,X-3 and $\sim 6$ times longer than in two other WR X-ray
binaries.

Thus, the ULX counterparts and SS\,433 occupy a region at this diagram between
O2--3.5If and WN5--7. This is also a region of ``intermediate temperature
LBV'' V\,532 and the ``LBV excursion'' of HD\,5980. The variability of the
He\,II lines of our counterparts in three consequent nights is shown by
the points connected by the lines. However, their behavior
in the He\,II diagram is nothing like stars. They exhibit night-to-night
variability both in the line width and equivalent width by a factor of
2--3. Variability in the radial velocity of the line is also
detected with amplitudes ranging from 100 to 400 km/sec \protect{\cite{fabr15}}.

If the ULX counterpart spectra were
produced from donor stars, variable surface gravity at about the same
phothospheric temperature would be required. Instead, the spectra may be
formed in unstable and variable winds formed in accretion disks. This
idea agrees with the fact that we do not find any regularities between
the EW, FWHM, and radial velocity of the He\,II line.

\begin{figure}
\hspace{2.0cm}
\includegraphics[width=.65\textwidth]{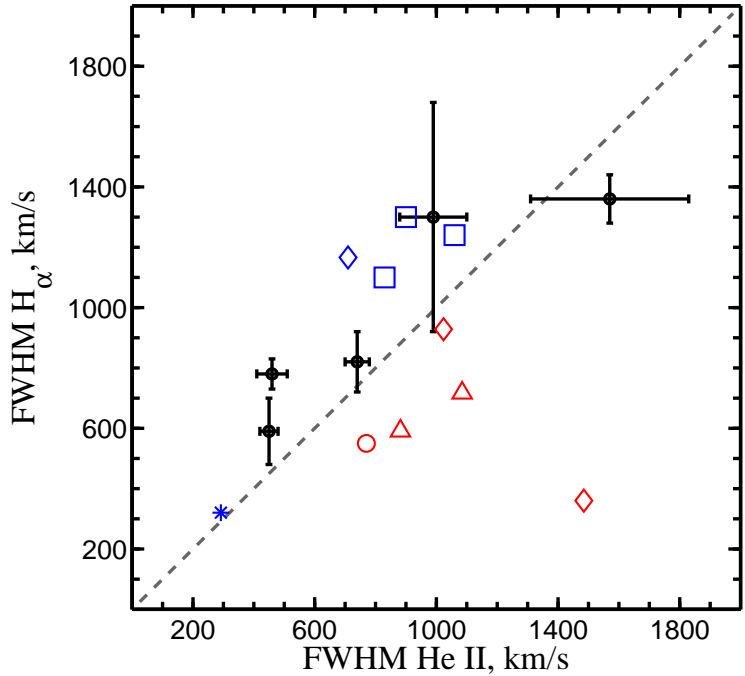}
\caption{Emission line widths of He II and H$\alpha$. The ULX counterparts with
simultaneous spectra (points with error bars) are shown from top to bottom:
NGC5204 X-1, Holmberg II X-1, NGC5408 X-1 (Cseh et al. 2013), NGC4559 X-7, 
Holmberg IX X-1. Also shown SS\,433 (blue diamond) from \protect{\cite{kubota10,Grandi1982}}
taken in the same precessional phase,
the LBV star V532 in M33 \protect{\cite{shol11}} in its hot state (asterisk), and 
three transitional stars (square) WR22/WN7ha, WR24/WN6ha, WR25/WN6ha \protect{\cite{Walborn2000}}. 
The transient black hole binaries with heated accterion disks look another: GX339-4 
\protect{\cite{Soria1999,Rahoui2014}} (circle), V404 Cyg \protect{\cite{Casares1991,Gotthelf1992}}
(triangle), and GRO J1655-40 \protect{\cite{Hunstead1997,Soria1998}} (red diamond). 
The X-ray transients may look as a probable analogue of IMBH with an accretion disk.
}
\label{fig5}
\end{figure}

Therefore we can exclude the case where these ULXs actually have WNL donors 
and their stellar winds produce the observed optical spectra.
The rapid variability of the He\,II line-width is difficult to be
explained, because the wind terminal velocity in stars is determined by
the surface gravity. Other arguments may be found in \protect{\cite{fabr15}}.
Opposite situation is in NGC\,300\,X-1,
M\,101\,ULX-1, and IC\,10\,X-1, where the He\,II line is stable and
has not a strong variability, this line is also shows a clear orbital
solution.

A clear difference between strong outflows and heated accretion disks is shown in
Fig.\,5. Indeed, such emission lines from the self-irradiated disk are observed
from Galactic stellar-mass black hole X-ray binaries in ourbursts, such
as GRO J1655--40, V404 Cyg and GX 339--4. In these cases, the width of
the He\,II lines is broader than hydrogen lines, as expected. 
We can interpret that their irradiated disks are not blocked by disk winds 
completely, and hence we observe both the He\,II and H$\alpha$ emissions directly 
at the disk surface. 

By contrast, our optical spectra of the ULXs have revealed 
that the He\,II emission line is always narrower than the H$\alpha$ (Fig.\,5)
line. This fact would be very difficult to be explained by standard
disks. Such a situation is observed in all hot supergiants with strong winds, 
WNL, LBV stars and SS\,433. When the wind is accelerated, upper and farter  parts 
of the wind are turning colder and have higher velocities. It is the best evidence
that  of a supercritical accretion in ULXs.

\begin{figure}
\hspace{2.2cm}
\includegraphics[width=.65\textwidth]{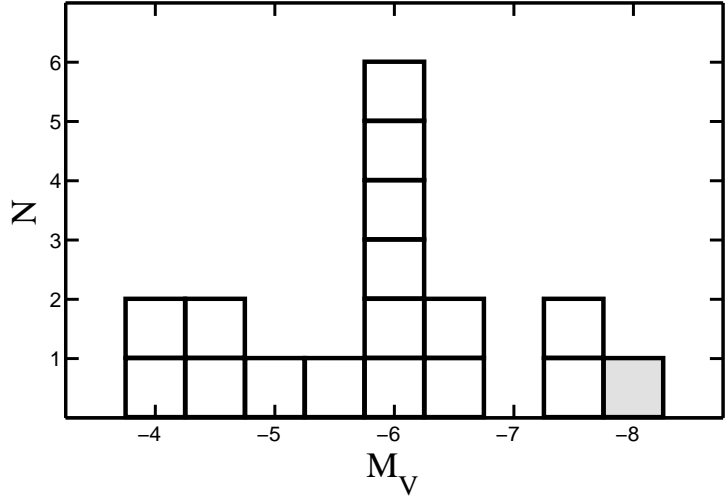}
\caption{Absolute magnitudes of all well-studied ULXs and SS\,433 (shadowed). The data are
from \protect{\cite{Tao2011,Vinokurov2013,Avdan2016,Vinokurov2016}}.     }
\label{fig6}
\end{figure}            
                       
In Fig.\,6 we show absolute magnitudes of the well-studied ULXs. The ULXs have 
a wide distribution with a well-defined maximum at $M_V \approx -6$. In decreasing 
optical luminosity they are SS\,433, NGC6946 ULX-1, NGC7793 P13, NGC4559 X-7, 
NGC5408 X-1, NGC5204 X-1, NGC4395 X-1, M81 ULS1, Holmberg II X-1, IC342 X-1, 
Holmberg IX X-1, NGC4559 X-10, NGC1313 X-2, NGC5474 X-1, NGC1313 X-1, M66 X-1 
and M81 X-6. The faintest ULX counterparts have optical luminosities of
25--26 magnitudes, they may be objectively detected with the Hubble Space Telescope. 
The abrupt decrease in the number of objects with decreasing $M_V$ probably could not 
be related to the effects of observational selection.

The total luminosity of a supercritical disk is proportional to the
Eddington luminosity with an additional logarithmical factor depending on
the original mass accretion rate \protect{\cite{SS73,pout07}}, because
the excess gas is expelled
as a disk wind and the accreted gas is advected with the photon
trapping, contributing little to the photon luminosity.
However, the UV and optical luminosity in such disks may strongly
depend on the original mass accretion rate, because these budgets are mainly
produced by the reprocess of the strong irradiation from the
wind (the excess gas). Therefore if the X-ray luminosity may be about 
the same, the optical luminosity may notably depends on the accretion rate.

It was found \protect{\cite{fabr15}} that the mass accretion rates in the ULXs
may be by a factor of 1.5--6 smaller and their wind temperatures are
by 1.4--4 times higher than those in SS\,433. Optical spectra of SS\,433 and the ULX
counterpart are nearly the same, but in X-rays they are drastically
different because we cannot observe the funnel in SS\,433 directly.

\section{Conclusions}

Why SS\,433 is not strong an X-ray emitter? We observe it nearly edge-on 
and we could not looke at its funnel \protect{\cite{Filippova2006}}. We observe only the 
X-rays jets and the illuminating radiation coming from deeper funnel regions
as reflected spectrum of the outer 
funnel walls \protect{\cite{Medvedev2010}}. In \protect{\cite{Khabibullin2016}} have been found 
constraints from SS\,433 funnel as a reflected signal in the Galactic plane 
(the molecular clouds). An upper limit was of $2 \times 10^{39}$~erg\,s$^{-1}$ 
in the 2-10 keV energy band. 

However, SS\,433 can not be a copy of ULXs because its optical luminosity is
notably higher than in ULX counterparts (Fig.\,6), and wind temperature is less
than in ULXs. In a study of optical filaments in W\,50 surrounding extended X-rays
jets, He\,II and hydrogen lines were discovered there. It was found using the ionization
thresholds of He\,II and H\,I in the filaments \protect{\cite{FabrikaShol2008}} that a funnel in SS\,433 
has a temperature $50 - 70$\,kK and its isotropic UV luminosity is $\sim 10^{40}$\,erg\,s$^{-1}$.
In such a case SS\,433 may be close relative to ULSs \protect{\cite{Urquhart2016}.
In well-studied ULSs, for example M81 ULS-1 and NGC247 ULS, their optical luminosities
are M$_V \approx -6.1$ and M$_V \approx -5.2$ respectively. The supercritical disk in SS\,433
has a luminosity of M$_V \approx -7.8$. The optical luminosity in SS\,433 
is bigger than in standard ULSs. 
 
We may conclude that the ULX/ULSs have about face-on supercritical
disks, nevertheless their optical luminosities and winds are not as powerful as
that in SS\,433. In this sense, SS\,433 remains the unique object.

\acknowledgments

The research was supported by the Russian Science Foundation grant (N\,14-50-00043)
and RFBR grant (N\,16-02-00567). 


\begin{thebibliography}{99}


\bibitem{feng11} H. Feng \& R. Soria {\em New Astron. Rev.} {\bf 55}, 166 (2011)

\bibitem{Bachetti2016} M. Bachetti {\em AN} {\bf 337}, 349 (2016)

\bibitem{madau01} P. Madau \& M.J. Rees {\em ApJ} {\bf 551}, L27 (2001)

\bibitem{fabr01} S. Fabrika \& A. Mescheryakov {\em IAUS} {\bf 205}, 268 (2001)

\bibitem{fabr1997} S.~N. Fabrika {\em Ap\&SS} {\bf 252}, 439 (1997)

\bibitem{fabr04} S. Fabrika {\em ASPRv} {\bf 12}, 1 (2004)

\bibitem{Bachetti2014} M. Bachetti, F. A. Harrison, D. J. Walton, et al. {\em Nature} {\bf 514}, 202 (2014)

\bibitem{Furst2017} F. F{\"u}rst, D.~J. Walton, D. Stern, et al. {\em ApJ} {\bf 834}, 77 (2017)

\bibitem{Israel2017} G.~L. Israel, A. Papitto, P. Esposito, L. Stella, et al. {\em MNRAS} {\bf 466}, L48 (2017)

\bibitem{Kawashima2016} T. Kawashima, S. Mineshige, K. Ohsuga \& T. Ogawa {\em PASJ} {\bf 68}, 83 (2016)

\bibitem{Stobbart2006} A.-M. Stobbart, T.~P. Roberts \& J. Wilms {\em MNRAS} {\bf 368}, 397 (2006)

\bibitem{gladst09} J.~C. Gladstone, T.~P. Roberts \& C. Done {\em MNRAS} {\bf 397}, 1836 (2009)

\bibitem{Sutton2013} A.~D. Sutton, T.~P. Roberts \& M.~J. Middleton {\em MNRAS} {\bf 435}, 1758 (2013)

\bibitem{Urquhart2016} R. Urquhart \& R. Soria {\em MNRAS} {\bf 456}, 1859 (2016)

\bibitem{Pinto2016a} C. Pinto, W. Alston, R. Soria, et al. {\em arXiv:1612.05569} (2016)

\bibitem{Fabrika2006} S. Fabrika, S. Karpov, P. Abolmasov \& O. Sholukhova {\em IAUS} {\bf 230}, 278 (2006)

\bibitem{Fabrika2007} S.~N. Fabrika, P.~K. Abolmasov \& S. Karpov {\em IAUS} {\bf 238}, 225 (2007)

\bibitem{Middleton2015} M.~J. Middleton, D.~J. Walton, A. Fabian et al. {\em MNRAS} {\bf 454}, 3134 (2015)

\bibitem{Pinto2016b} C. Pinto, M.~J. Middleton \& A.~C. Fabian {\em Nature} {\bf 533}, 64 (2016)

\bibitem{Soria2016} R. Soria \& A. Kong {\em MNRAS} {\bf 456}, 1837 (2016)

\bibitem{abol07} P. Abolmasov, S. Fabrika, O. Sholukhova \& V. Afanasiev
{\em Astrophys. Bull.} {\bf 62}, 36 (2007)

\bibitem{Grise2012} F. Gris{\'e}, P. Kaaret, S. Corbel, et al. {\em ApJ} {\bf 745}, 123 (2012)

\bibitem{Cseh2012} D. Cseh, S. Corbel, P. Kaaret, et al. {\em ApJ} {\bf 749}, 17 (2012)

\bibitem{Lehmann2005} I. Lehmann, T. Becker, S. Fabrika, et al. {\em A\&A} {\bf 431}, 847 (2005)

\bibitem{Egorov2017} O.~V. Egorov, T.~A. Lozinskaya \& A.~V. Moiseev {\em MNRAS} {\bf 467}, L1 (2017)

\bibitem{pout13} J. Poutanen, S. Fabrika, A. Valeev, et al. {\em MNRAS} {\bf 432}, 506 (2013)

\bibitem{fabr15} S. Fabrika, Y. Ueda, F. Vinokurov, O. Sholukhova \& M. Shidatsu {\em Nat. Phys.} {\bf 11}, 551 (2015).

\bibitem{kubota10} K. Kubota, Y. Ueda, S. Fabrika, et al. {\em ApJ} {\bf 709}, 1376 (2011)

\bibitem{shol11} O.~N. Sholukhova, S.~N. Fabrika, A.~V. Zharova, et al. {\em Astrophys. Bull.} {\bf 66}, 123 (2011)

\bibitem{crowther11} P.~A. Crowther \& N.R. Walborn {\em MNRAS} {\bf 416}, 1311 (2011)

\bibitem{Motch2014} C. Motch, M.~W. Pakull, R. Soria, et al. {\em Nature} {\bf 514}, 198 (2014)

\bibitem{Liu2015} J.-F. Liu, Y. Bai, S. Wang, S. Justham, et al. {\em Nature} {\bf 528}, 108 (2015)

\bibitem{Crowther1997} P.~A. Crowther \& L.~J. Smith {\em A\&A} {\bf 320}, 500 (1997)

\bibitem{roberts11} T.~P. Roberts, J.~C. Gladstone, A.~D. Goulding, et al. {\it AN} {\bf 332}, 398 (2011)

\bibitem{Cseh2011} D. Cseh, F. Gris{\'e}, S. Corbel \& P. Kaaret {\it ApJ} {\bf 728}, L5 (2011)

\bibitem{Crowther2010} P.~A. Crowther, R. Barnard, S. Carpano, et al. {\em MNRAS} {\bf 403}, L41 (2010)

\bibitem{kuntz05} K.~D. Kuntz, R.~A. Gruendl, Y.-H. Chu, et al. {\em ApJ} {\bf 620}, L31 (2005)

\bibitem{silverman08} J.~M. Silverman \& A.~V. Filippenko {\it ApJ} {\bf 678}, L17 (2008)

\bibitem{vanKekrwijk96}
M.~H. van Kerkwijk, T.~R. Geballe, D.~L. King, et al. {\it A\&A} {\bf 314}, 521 (1996)


\bibitem{Liu2013} J.-F. Liu, J.~N. Bregman, Y. Bai, et al. {\em Nature} {\bf 503}, 500 (2013)

\bibitem{Grandi1982} S.~A. Grandi \& R.~P.~S. Stone {\em PASP} {\bf 94}, 80 (1982)

\bibitem{Walborn2000} N.~R. Walborn \& E.~L. Fitzpatrick {\em PASP} {\bf 112}, 50 (2000)

\bibitem{Soria1999} R. Soria, K. Wu \& H.~M. Johnston {\em MNRAS} {\bf 310}, 71 (1999)

\bibitem{Rahoui2014} F. Rahoui, M. Coriat \& J.~C Lee {\em MNRAS} {\bf 442}, 1620 (2014)

\bibitem{Casares1991} J. Casares, P.~A. Charles, D.~H.~P. Jones, et al. {\em MNRAS} {\bf 250}, 712 (1991)

\bibitem{Gotthelf1992} E. Gotthelf, J.~P. Halpern, J. Patterson \&  R.~M. Rich {\em AJ} {\bf 103}, 219 (1992)

\bibitem{Hunstead1997} R.~W. Hunstead, K. Wu \& D. Campbell-Wilson {\em ASPC} {\bf 121}, 63 (1997)

\bibitem{Soria1998} R. Soria, D.~T. Wickramasinghe, R.~W. Hunstead \& K. Wu {\it ApJ} {\bf 495}, L95 (1998)

\bibitem{Tao2011} L. Tao, H. Feng, F. Gris{\'e} \& P. Kaaret {\it ApJ} {\bf 737}, 81 (2011)

\bibitem{Vinokurov2013} A. Vinokurov, S. Fabrika \& K. Atapin {\em Astrophys. Bull.} {\bf 68}, 139 (2013)

\bibitem{Avdan2016} S. Avdan, A. Vinokurov, S. Fabrika, et al. {\em MNRAS} {\bf 455}, L91 (2016)

\bibitem{Vinokurov2016} A. Vinokurov, S. Fabrika \& K. Atapin {\em arXiv:1606.03024} (2016)

\bibitem{SS73} N.~I. Shakura \& R.~A. Sunyaev {\em A\&A} {\bf 24}, 337 (1973)

%
\bibitem{pout07} J. Poutanen, G. Lipunova, S. Fabrika, et al. {\em MNRAS} {\bf 377}, 1187 (2007)

\bibitem{Filippova2006} E. Filippova, M. Revnivtsev, S. Fabrika, et al. {\em A\&A} {\bf 460}, 125 (2006)

\bibitem{Medvedev2010} A. Medvedev S. \& Fabrika {\em MNRAS} {\bf 402}, 479 (2010)

\bibitem{Khabibullin2016} I. Khabibullin \& S. Sazonov {\em MNRAS} {\bf 457}, 3963 (2016)

\bibitem{FabrikaShol2008} S.~N. Fabrika \& Sholukhova {\em Proceedings of the VII Microquasar Workshop: Microquasars and Beyond} p. 52 (2008)
 
\end{thebibliography}
\end{document}